\documentclass{article}
\usepackage{varioref}
\usepackage{amssymb}
\usepackage{color}

\begin{document}


\title{\bf  Time evolution of the inside  of the black hole's horizon}
\author{H. Hadi\thanks{%
email: hamedhadi1388@gmail.com} ~ and ~ F. Darabi\thanks{email: f.darabi@azaruniv.ac.ir%
}
\\{\small Department of Physics, Azarbaijan Shahid Madani University, Tabriz,
Iran}}
\date{\today} \maketitle
\begin{abstract}
We consider the Wheeler-DeWitt equation near the horizon of the black hole
where the entangled vacuum state  is chosen as the static universe
state. Then, using the entangled property of the vacuum state, we investigate
the dynamical evolution of the subsystems, namely inside and outside of the horizon. \end{abstract}
\section{Introduction}

It is well known that the canonical quantization of general relativity yields the Wheeler-DeWitt equation $\cite{dew,har}$. This equation leads to static state of the universe
as well as  the problem of time $\cite{kuch,ish,ash,and,mer,marletto,barbour}$.
To overcome this problem a solution was suggested by Page and Wootters (PaW)
$\cite{pag,woo}$. By considering quantum entanglement, a static system can
be described as an evolving universe by the view of internal observers. An
hypothetical external observer may describe clock system and the rest of the universe as a whole system in a stationary state. This system will be evolving from the view of internal observers that test correlation between the clock and the rest $\cite{pag,woo,pagg,gam,per,rov}$. Thus, entanglement
between subsystems provide the possibility to describe time as an emergent property
of the subsystems of the universe. For an experimental illustration refer
to $\cite{more}.$ In this paper, we apply PaW mechanism to
the near horizon of the black hole to study the time evolution of the black
hole's interior. We investigate this mechanism within two different paradigms, $ER=EPR$ $\cite{mal}$ and
firewalls $\cite{amps}$.

The complementarity view of black holes has been threatened by firewall
concept. $ER=EPR$ conjecture in preserving complementarity $\cite{mal}$ does not comply
with AMPS which proposes firewall at the horizon of black hole to avoid APMS's paradox $\cite{amps}$ . AMPS has  argued that considering the complementarity there is a contradiction
in accepting all three following assumptions at once: 1) an evaporating black hole preserves quantum information without destroying it (unitarity), 2) the event horizon of black hole is not unusual for an in-falling observer crossing it, 3) an observer staying outside the black hole  works with relativistic effective quantum field theory. AMPS considers the late radiation $B$ of an old black hole (emitted half of its radiation away \cite{old})
 as maximally entangled with its early radiation $R_{B}$. Assumptions 1 and 3 require the $B$ to be entangled with a subsystem of $R$, and on the other hand, the assumption 2 leads to entanglement between $B$ and a subsystem of interior of the black hole. This violates the monogamy of quantum entanglement
\cite{mon1,mon2}. It
asserts that if two quantum systems are maximally entangled, non of them can
be entangled with a third system. To overcome this puzzle, AMPS argue that
there is only one singularity at firewall and no interior of black hole exists
\cite{amps,fire11}.

One of the solutions to overcome AMPS's paradox without violation of equivalence
principle near the horizon is the $ER=EPR$ conjecture.
The
$ER$ bridge from one hand and $EPR$ pair on the other hand have
a relation by $ER=EPR$ \cite{mal}. This means that $ER$ 
bridge is created by $EPR$ correlation in the microstates of two entangled
black holes.
This result is based on the works \cite{B1},\cite{B2}. To explain more, the
$EPR$ correlated quantum system is nothing but a weakly coupled Einstein gravity description. In other words, the $ER$ bridge is a highly quantum
object. There are some speculations that for every singlet state there exists a quantum bridge of this type. For more discussion of AMPS's paradox and
another solution for it, refer to $\cite{preskill}$.

In this paper, we study the black hole's near horizon features and PaW mechanism
briefly in section 2. In the third section, the dynamical evolution of the
black hole's interior is studied within $ER=EPR$ paradigm using Wheeler-DeWitt equation near the horizon of the black hole. This is repeated in the section 4, concerning firewall at the horizon of black hole. At the end we have
a conclusion section.


\section{Black hole's near horizon features and PaW mechanism}
In the black hole formation and evaporation process, the unitarity of S-matrix
is an important fact. We assume that $B$ is an outgoing Hawking mode in
the near horizon zone of a black hole. The unitarity of S-matrix imposes that the mode $B$
at near horizon be pure for a newly constructed black hole, otherwise, it has to be purified as 
a whole Hawking radiation, emitted partly at near horizon, entangled with
the other part at far
distance,
for an old black hole.  In the later case, the exact purification of the  $B$ mode is associated to
degrees of freedom of the black hole.

Before considering the evaporation and radiation of an old black hole\footnote{Old black hole is known as a black hole which has radiated more than half of its initial entropy in the Page time. One
can consider an old black hole by the collapse of some pure state and then
evaporation of it into Hawking radiation which can be divided into early  and
late parts as follows
$|\Psi>= \sum_{i}|\psi_{i}>_{E}\otimes |i>_{L}$. }, the entropy of which  is smaller than the entropy of the radiation  that it has
already emitted, there is a so called AMPS paradox. Here the entropy means von Neumann entropy. This entropy can be written for two quantum systems as follows
\begin{equation}
	S_{AB}=-tr(\rho_{AB} ln\rho_{AB})
\end{equation}
where $\rho_{AB}$ is density matrix for quantum mechanical systems $A$ and $B$. The amount of von Neumann entropy $S_{A}$ ($S_{B}$) is considered by
\begin{equation}
	S_{A}=-tr(\rho_{A} ln\rho_{A}) 
\end{equation}
\begin{equation}
	S_{B}=-tr(\rho_{B} ln\rho_{B}) 
\end{equation}
which is derived by tracing over states $B$ ($A$) in density matrix $\rho_{AB}$. 
 When $A$ and $B$ are maximally entangled (not pure) then $S_{A}=S_{B}=1$ and $S_{AB}=0$. On the other hand when $S_{A}=S_{B}=0$, then there is no any entanglement between $A$ and $B$ (pure).  

To consider the AMPS paradox let's do as follows. For an exterior rest observer,  the outgoing near horizon Hawking mode $B$ has
the entropy $S_{B}\simeq 1$ which indicates that it is not pure. However, this mode can be purified  by the early emitted Hawking radiation.
If we denote $R_{B}$ for the early radiation, then the von Neumann entropy
$S_{BR_{B}}$ is exponentially small, namely $S_{BR_{B}}\ll1$. If we indicate the interior mode of  black hole by $A$, then for an in-falling observer,  realizing
the vacuum,  the mode $B$ has to be purified by $A$. In other
words, $S_{BA}\ll1$. On the other hand the sub-additivity theorem implies\cite{fire11}
\begin{equation}
	S_{B}\leqslant
	S_{BA}+S_{BR_{B}}
\end{equation}
 which is violated by the simultaneous
imposition of  the   results $S_{BR_{B}}\ll1$ and $S_{BA}\ll1$.
Thus, in order to revalidate this theorem, the statement of {\it entanglement
monogamy} is introduced, which allows each state to be entangled with one
and only one other state \cite{mon2}.  

To overcome above paradox, AMPS suggested the existence of firewall at the
horizon which is created by breaking of entanglement between $B$ and $A$. 
The monogamy of entanglement does not allow entanglement among three
parties. In AMPS's suggestion one of the entanglements breaks down which leads to
the creation of firewall at horizon. This violates the equivalence principle
of general relativity near the horizon. 

Regarding  these properties of the black hole which leads to "frozen vacuum"
$\cite{bosso}$, in the next section we will consider the Wheeler-DeWitt equation in the near horizon
of black hole to ascribe a typical time evolution for the quantum states, 
inside the black hole.

 Before describing our argument, we review PaW approach which
is necessary for our discussion, as follows.   
\begin{itemize}
\item 
The universe is timeless
\begin{equation}\label{wdequation}
H|\psi>=0,
\end{equation}
where $|\psi>  \in  \mathcal{H}$  is an eigenstate of its Hamiltonian $H$.
\end{itemize}

\begin{itemize}
\item 
Hamiltonian includes at least one good clock. It means that a clock system
$H_{c}$
with a large distinguishable states, interacts weakly (or does not have interaction at all) with the rest of universe $H_{r}$. So, it leads to Hamiltonian system
with tensor product structure in its eigenstate space $\mathcal{H} \in \mathcal{H}_{c}\otimes \mathcal{H}_{r}$
such that non-interacting property holds:
\begin{equation}\label{H}
H=H_{c}\otimes I_{r}+I_{c}\otimes H_{r},
\end{equation}
where $I$ are the unit operator on each subsystem.
\end{itemize}  
\begin{itemize}
\item 
Clock and the rest of the universe are entangled. This feature allows the
apparent dynamical evolution of  the rest of  universe in terms of clock,
without any evolution
at the level of the universe, at all.

To explain it in more details, one assumes that the state of the universe
is $|\psi>$, then $|\psi (t)>_{c}$ and $|\psi (t)>_{r}$ are the states of clock system and rest of the universe system, respectively. By projecting $|\psi>$
on the states of clocks $|\psi (t)>_{c}$, and considering $|\psi (t)>_{c}= e^{-iH_{c}t/\hslash}|\psi (0)>_{c}$, one gets the vectors
 \begin{equation}\label{pi}
 |\psi (t)>_{r}:=_{c}<\psi(t)|\psi>=  e^{-iH_{r}t/\hslash}|\psi (0)>_{r}.
 \end{equation}
 This indicate the proper evolution of subsystem $r$ under the action of its
 local Hamiltonian $H_{r}$. Although the system globally appears to be static,
 its subsystems indicate correlations which represent an apparent dynamical evolution. In fact, this is called evolution without evolution.   
\end{itemize}

\section{Time evolution of the interior of a black hole  and $ER=EPR$}
In this section, we investigate the time evolution of an old black hole's interior
according to Wheeler-DeWitt equation by considering $ER=EPR$ conjecture. The left hand side of the $ER=EPR$ is Einstein Rosen bridge and the right hand side of it is the $EPR$ paradox. There are similarities between entanglement pair $EPR$ and Einstein Rosen bridge. To show that, suppose a large number of particles which are separated in a two entangled Bell pairs. Each part is collapsed to make a single black hole. Now there are two entangled black holes which can be connected by $ER$ bridge. In other words two entangled black holes ($EPR$ pairs) can do the role of $ER$ bridge. It is important to mention that this relation is over a particular manifold and maybe it cannot be applied in every spacetime. However, some physicist take a radical position that these two parts are linked even for a single entangled pair $\cite{mal}$. For our goal in this section and whole of this paper, since we have constrained our consideration to black holes, the symbolic equation $ER=EPR$ is applicable. It is important to note that both side of the $ER=EPR$ conjecture, have "no superluminal signals" and "no creation by LOCC" features. There is no violation of locality in the both entanglement and Einstein Rosen bridge part of the equation or in other words there is no superluminal signals in $ER$ bridge and $EPR$ Bell pairs. The other feature, no creation by LOCC, admits that by local operation and classical communication (LOCC) one can not increase or create the entanglement of both parts. In other words, Alice with her entangled pair by doing local operation and sending information by classical communication cannot create or increase the entanglement. The same situation also true for Einstein Rosen bridge part. For two distant black hole with no Einstein-Rosen bridge, there does not seem to be any way to create a bridge between them without preexisting bridges.

One of the applications of $ER=EPR$ conjecture is to overcome the AMPS paradox. For an old black hole the interior a exterior states of the near horizon is indicated by $A$ and $B$. The early Hawking radiation's state is $R_{B}$. The states $B$ and $A$ are entangled and on the other hand the states $B$ are also entangled by early Hawking radiation $R_{B}$. As we mentioned in the previous sections $B$ cannot be entangled both with $A$ and $R_{B}$ (monogamy of entanglement). To overcome this paradox, $ER=EPR$ can be applied here by mapping interior states $A$ by $ER$ bridge to early Hawking radiation $R_{B}$. Therefore the monogamy of entanglement is not violated because the interior states $A$ and early Hawking radiation are identified$(A=R_{B})$.

With these descriptions, horizon of the black hole for an in-falling observer is not a particular region and he/she can cross the horizon without experiencing any particular event(without confronting with firewall). However, by applying $ER=EPR$ not only for quantum vacuum $A$ and $B$ states and Hawking radiation $R_{B}$, but for the exited states of the vacuum, we will confront with a particular vacuum near the horizon of the black hole which is called frozen vacuum. To understand the essence of this vacuum and its relation to $ER=EPR$ conjecture, suppose two observers an in-falling observer Alice and a static observer Bob for an old black hole. We indicate the Hilbert state of $A$, $B$ and $R_{B}$ states, by $|n>_{\tilde b}$, $|n>_{b}$ and $|n>_{R_{B}}$, respectively.  

Now we want to consider the vacuum when it is exited and to observe its influence in  $ER=EPR$ paradigm. In doing so, suppose an old black hole and indicate a thermally entangled state without normalization factors of $bR_{B}$ as follows
\begin{equation}\label{pointerpremeasurement}
	|\psi>_{pR_{B}b}=|i>_{p}\otimes\sum_{n=0}^{\infty}|n>_{b}|n>_{R_{B}}
\end{equation}
Here $|i>_{p}$ is the state of a pointer which has not interacted with any of the subsystems yet. By using $ER=EPR$ conjecture which here is $A=R_{B}$ one can apply the following map
 \begin{equation}
 |0>_{R_{B}}\rightarrow |0>_{\tilde b},...
 |j>_{R_{B}}\rightarrow |j>_{\tilde b},...   
 \end{equation}
If we assume the black hole is billions of light years, then the curvature is negligible in the near horizon region. In this vicinity 
one may expect the violation of semiclassical approach or equivalence principle.  

To complete premeasurement we suppose the pointer $p$ measures the states $R_{B}$. So the equation (\ref{pointerpremeasurement}) becomes 
\begin{equation}\label{pre2}
	|\psi>_{pR_{B}b}=\sum_{n=0}^{\infty}|n>_{b}|n>_{R_{B}}|n>_{p}
\end{equation} 
A realistic system cannot be separated from environment. Then, here a pointer can do the role of the environment for radiation states 
$R_{B}$. If one trace over the stats $B$ the rest $pR_{B}$ is a mixed state and is not pure. Therefore, any map from $R_{B}$ to states $A$ cannot give the vacumm state of the near-horizon zone $|0>_{b\tilde b}$. So, if we include the environment $p$ for Hawking radiation states $R_{B}$ the donkey map becomes
 \begin{equation}
|0>_{R_{B}}|0>_{p}\rightarrow |0>_{\tilde b},...
|j>_{R_{B}}|j>_{p}\rightarrow |j>_{\tilde b},...   
\end{equation} 
The in-falling vacuum is proportional to
\begin{equation}\label{vacuumapp}
	|0>_{b\tilde b} \alpha\sum_{n=0}^{\infty}x^{n}|n>_{b}|n>_{\tilde b},
\end{equation}
where we suppressed the normalization factor.  

Now suppose the pointer $p$ measures $b$ instead of $R_{B}$. This gives the same results as equation (\ref{pre2}) . In addition, assume Bob an static observer who is one light year from near-horizon zone is aware of this measurement then he disappears. Nine years later a clueless Alice who is a free falling observer is going to experience the near horizon vacuum. In her journey she will not recognize any thing especial near the horizon vacuum because from her knowledge of black hole she know that near horizon vacuum $B$ can be purified by states $A$ which is identified by $R_{B}$. Alice was aware of $R_{B}$ before starting her journey into vacuum and then she was aware of $A$, too. Therefor she enjoy her journey and will not see anything except the in-falling vacuum.       

However, if Alice become aware of Bob's knowledge she will confront with a contradiction. In other words if Bob meet the Alice and share the $p$ measurement of vacuum $B$, then Alice in purifying of $A(=R_{B})$ with $B$ confronts with a contradiction. Because, in this situation $B$ is not purified by $A$ from Alice's view. To avoid this contradiction Alice must always experience the in-falling vacuum (\ref{vacuumapp}) and she cannot see any exited vacuum by the pointers, environment or even by herself. Then, near-horizon vacuum is an special vacuum which is called "Frozen Vacuum".    

As we mentioned before we want to construct wheeler-DeWitt equation in near-horizon zone. We recognized that near-horizon zone is frozen vacuum. To construct the wheeler-DeWitt equation in this vicinity we use the page and wootter approach $\cite{pag}$ We reviewed this approach in section (2). Now its time to apply wheeler-DeWitt to near-horizon zone. In doing so, we start from vacuum state of near horizon or frozen vacuum.

According to Bousso, the in-falling vacuum state without normalization factors
is as follows $\cite{bosso}$
\begin{equation}
|0>_{b\tilde{b}}=\sum_{n=0}^{\infty}x^{n}|n>_{b}|n>_{\tilde b},
\end{equation}
where $|n>_{b}$ and $|n>_{\tilde b}$ are the quantum states of outside and inside
the black hole horizon, from in-falling observer's point of view, and the coefficient $x=e^{-\beta \omega/2}$ for modes with Killing frequency of the order of Hawking temperature is of order one. This is particular vacuum state which is called "frozen vacuum". The observer in this vacuum state, near the horizon, is unable to observe any particle,
whereas a rest inertial observer far from gravity is able to observe particles from
her/his vacuum state. In other words, it leads to violation of equivalence principle. This vacuum state
is the only state that exists near the horizon  when one is in the $ER=EPR$ paradigm. It turns out that while $ER=EPR$ conjecture   tries to save monogamy principle in black hole
physics, 
 at the same time leads to violation of equivalence  principle (through the frozen vacuum rather than the firewall). These explanations
have far-reaching implication for our next arguments.    

Now, we consider the frozen vacuum state as the universe state. Since there is only
 one vacuum state - frozen vacuum state - near the horizon  in the $ER=EPR$
 case, it is the mere state that can be described as the universe
state. The  local Hamiltonians for subsystems $c$ and $r$, defined by relation ($\ref{H}$), are given
by  $H_{b}$ and
$H_{\tilde b}$, respectively as  
\begin{equation}\label{Hb}
 H_{b}=\sum_{n=0}^{\infty}x^{-n}|n>_{bb}<n|,
 \end{equation}
  \begin{equation}\label{Hbb}
 H_{\tilde b}=-\sum_{n=0}^{\infty}x^{-n}|n>_{\tilde b\tilde b}<n|,
 \end{equation}
 where $H_{b}$ and $H_{\tilde b}$ indicate the local Hamiltonians for outside (clock system) and inside (rest of universe)
the black hole horizon, from in-falling observer's point of view, respectively. Now by using equations (\ref{wdequation}), (\ref{H}),
	(\ref{Hb}) and (\ref{Hbb}) one can obtain the following equation
\begin{equation}
	\left(\sum_{n=0}^{\infty}x^{-n}|n>_{bb}<n|\otimes I_{\tilde{b}} -I_{b}\otimes\sum_{n=0}^{\infty}x^{-n}|n>_{\tilde b\tilde b}<n|\right)|0>_{b\tilde{b}}=0
\end{equation} 
where we apply the vacuum state $|0>_{b\tilde{b}}$ as universe state This is wheeler-DeWitt equation for this model of system. Note that, as a whole, the constraint $H|\psi>=0$ is compatible with current approaches to quantum gravity. In other word, it can be interpreted as wheeler-DeWitt equation in a closed universe $\cite{dew}$. However, it can also be regarded as the first set of sufficient conditions for a timeless approach to time in quantum gravity.

 Now, the in-falling observer has equipped with Hamiltonian $H_{b}$ and
also knows the universe state by his knowledge of black hole's theory. This
knowledge includes $ER=EPR$ conjecture which identifies the interior $A$ of the black hole with the outside distant Hawking
radiation $R_{B}$, ($A=R_{B}$ which is called donkey map). This map also includes the interaction of  $R_{B}$ with anything outside, even the observer itself. Whatever happens to the Hawking
radiation $R_{B}$, the frozen vacuum for the in-falling observer does not
change and so this observer is still unable to observe any particle. For example, the observer can read the Hawking radiation $R_{B}$
and then  use donkey map  as follows
\begin{equation}
|n>_{R_{B}}\rightarrow |n>_{\tilde b},   
\end{equation}
for $n=0, 1, 2, 3, ...$ as quantum states. Therefore, the observer by the
knowledge
of $|n>_{R_{B}}$ can recognize $|n>_{\tilde b}$, and so construct
the frozen vacuum state $|0>_{b\tilde b}$ without falling into the interior of the black hole. For more
discussion refer to  $\cite{bosso}$. 

 To know the proper time evolution of the interior of the black hole by the PaW approach,   without  
falling into it, the exterior observer can use her/his own subsystem state
\begin{equation}
|\psi (t)>_{b}=e^{-iH_{b}t/\hslash}|\psi (0)>_{ b},
\end{equation}
where $|\psi (0)>_{ b}$ is the initial state of the subsystem $b$, and then  uses equation
($\ref{pi}$) to derive the proper time evolution of the interior of the black
hole, without falling into it, as follows   
\begin{equation}
|\psi (t)>_{\tilde b}:=_{b}<\psi(t)|0>_{b\tilde b}=  e^{-iH_{\tilde b}t/\hslash}|\psi (0)>_{\tilde b},
\end{equation} 
where $|\psi (0)>_{\tilde b}=_{ b}<\psi(0)|0>_{b\tilde b}$ is the initial
state of the subsystem $\tilde b$.

  We conclude that the observer who has access to the Hawking radiation $R_{B}$, has access to the internal of the black hole, too, without falling into
it. Therefore, he can also access to the Hamiltonian ($\ref{Hbb}$) near the horizon
outside of it. With these interpretations, he has ability to make a measurement
globally through $H$ because of his simultaneous accessibility to the Hamiltonian $H_{b}$ and $H_{\tilde b}$. The observer
 by considering
the whole system will recognize it as a static system, but by considering its disjoint
subsystems as clock-rest system,  will recognize it as a
dynamical system.
\section{Time evolution of the interior of  a black hole  and firewalls}
In this section, we investigate the time evolution of the interior of the
black hole in the presence of firewall that AMPS has  suggested for solving
the AMPS's paradox. Therefore, we study a little more about the firewall.
\subsection{Firewall}
AMPS has argued that for a black hole which has radiated more than half of
its initial entropy in the Page time, the firewall is created at the horizon
where the in-falling observer burns up there $\cite{amps}$. This is in
contradiction with both the equivalence
principle and  the postulate of black hole complementarity $\cite{leny}$.
AMPS claims that the firewall is constructed in scrambling time which is much less than Page time. However, in a more gradual picture of forming firewall in $\cite{leny}$, this is not a correct picture. For more explanation, consider
an old black hole with  early Hawking radiation $R$, the outside of the horizon
$B$ and the interior of the black hole $A$. For an old black hole, $B$ has entanglement
with Hawking radiation $R_{B}$. On the other hand, for in-falling observer
the interior $A$ and the outside $B$ are entangled. Now, suppose that Alice as in-falling observer, measures
the state of the $R_{B}$ and then falls into the black hole. She
has recognized the state of $R_{B}$ and in her journey into the black hole,
can measure
the state of $B$. As long as $B$ has entanglement with $R_{B}$, regarding the monogamy of entanglement, she must not recognize the entanglement between $B$ and $A$. To overcome this paradox, APMS argues that the entanglement between $A$ and $B$ breaks down for Alice. This
leads to firewall at horizon in scrambling time. According to $\cite{leny}$,  this is not a correct picture,
because the high degree of entanglement between $B$ and $R_{B}$ does not occur suddenly. The firewall is not a part of horizon but it is only as an extension of singularity.
The separation of the singularity from horizon is a gradual function of time
and at the Page time this separation goes to zero. In this time there is
no horizon at all and the singularity of  black hole is located at the location
of the horizon. So, an
in-falling observer terminates at horizon (singularity of black hole).
The story is different for the young black hole. In the case of young and large black holes the in-falling observer survives passing through the horizon.       
\subsection{Time evolution of the inside of the black hole}
Now, we investigate the time evolution of the black hole's inside from the
viewpoint of an in-falling observer outside the black hole, near the horizon in the presence
of firewall. According to all of above considerations about PaW approach
 in section 2, we choose the state of near horizon vacuum state as the universe state 
 \begin{equation}\label{a1}
|\psi>_{b\tilde b}=\frac{1}{\sqrt{2}}(|1>_{b}|1>_{\tilde b}+|0>_{b}|0>_{\tilde b}),
\end{equation}
which is identified by imposing the Wheeler-DeWitt equation as $H|\psi>_{b\tilde
b}=0$.

Now, we need local Hamiltonians for subsystems $c$ and $r$ which are $H_{b}$ and
$H_{\tilde b}$ respectively, as clock subsystem and the rest of the universe  obeying relation ($\ref{H}$) and are given by 

\begin{equation}\label{a2}
H_{b}=|1>_{b}<0|_{b}-|0>_{b}<1|_{b},
\end{equation}
\begin{equation}\label{a3}
H_{\tilde b}=|1>_{\tilde b}<0|_{\tilde b}-|0>_{\tilde b}<1|_{\tilde b},
\end{equation}
where $H_{b}$ indicates the local Hamiltonian for outside of observer near
horizon and $H_{\tilde b}$ for the interior region of horizon.

The observer is equipped with local Hamiltonian ($\ref{a2}$) near the horizon of
black hole. For a young and large black hole the in-falling observer without any concern
of existence of the firewall can measure the proper time evolution of the
black hole's interior. In doing so, what she needs is to do the following measurement \begin{equation}\label{fire}
|\psi (t)>_{\tilde b}:=_{b}<\psi(t)|\psi>_{b\tilde b}=  e^{-iH_{\tilde b}t/\hslash}|\psi (0)>_{\tilde b},
\end{equation}
where $|\psi (0)>_{\tilde b}=_{b}<\psi(0)|0>_{b\tilde b}$ and  $|\psi (t)>_{b}=e^{-iH_{b}t/\hslash}|\psi (0)>_{b}$ and  we know that $|\psi (0)>_{b}$ is the initial state of the subsystem $b$. Therefore, the correlation between
$_{b}<\psi(t)|$ and universe state $|\psi>_{b\tilde b}$ which comes from
entanglement between them provide the possibility for observer to measure proper
time evolution of the black hole's interior.

In the case of an old black hole, the observer again is equipped with local Hamiltonian ($\ref{a2}$) near the horizon of
the black hole. If the in-falling observer does not make any measurement on early
Hawking radiation $R_{B}$ there will be no detectable difference between young
and old black hole for her and she will not encounter any firewall at the
horizon. Therefore, she is able to measure the evolution of the subsystem $\tilde
b$ by the equation ($\ref{fire}$) using the correlation between subsystems
that mimic the presence of dynamical evolution. On the other hand, suppose
that the in-falling observer at first makes a measurement on early
Hawking radiation and then  near the horizon she makes a measurement on $B$. If
she recognizes $R_{B}$ and $B$ as maximally entangled, then she will confront with firewall
at the horizon which comes from the breaking down of entanglement between
$A$ and $B$. Therefore, in the lack of entanglement between $A$ and $B$ she will not be able to measure the dynamical evolution of the subsystem $\tilde b$ by equation
($\ref{fire}$). In other words, we can conclude that she does not recognize
any evolution inside the black hole. This conclusion is very close to the
approaches  claiming that the lack of entanglement between two sides of horizon leads to non existence of the entire space-time behind the firewall $\cite{non1,
non2, non3}$.

\section{Conclusion}
Although there is a frozen vacuum near the horizon region, one can construct the
Wheeler-DeWitt equation there and study the dynamical evolution of the system.
If one accepts the $ER=EPR$ conjecture, then  the time evolution of
the interior of the horizon can be accessed by an infalling observer before
crossing the horizon. The outside observer of the black hole  can do measurement on early Hawking radiation and then by  the help of $ER=EPR$ conjecture and the map $A=R_{B}$ (donkey map)  can  access to the interior states.
Next, the observer can construct Wheeler-DeWitt operator (Hamiltonian) near the horizon to 
 operate on the frozen vacuum as the universe state with zero energy,
and determine the local Hamiltonians for outside and inside
the black hole horizon. Finally, the observer is able to obtain the time evolution of the interior states of the black hole by  using the outside subsystem and the  frozen vacuum state. 

If  the observer be in the firewall paradigm, she/he will confront with two cases:
For young black hole, the observer is equipped with his local Hamiltonian  near the horizon of
black hole. For a young and large black hole, the in-falling observer without any  concern about the existence of  
firewall can describe the proper time evolution of the
black hole's interior. 

In the case of old black hole, if the observer does not make any observation on the early Hawking radiation, she/he cannot distinguish between old and young black holes, and so  repeats the same calculation of young black hole for the old one. But if the observer makes observation on the  early Hawking radiation, then she/he will confront with a firewall and there is no any time evolution on the other side of the horizon.

  \section{Acknowledgments}

 This work is based upon research funded by Iran National Science Foundation (INSF) under project No 99033073.  



\begin{thebibliography}{99}

\bibitem{dew}B. S. DeWitt, Quantum theory of gravity, Phys. Rev.160, 1113(1967).
\bibitem{har}J.B. Hartle, S. W. Hawking, Wave function of the Universe Phys, Rev. D 28, 2960-2975 (1983).
\bibitem{kuch}K.V. Kuchar, In Proceedings of the 4th Canadian Conference
on General Relativity and Relativistic Astrophysics"
ed. G. Kunstatter, D. Vincent and J. Williams
7 (World Scientific, Singapore 1992).
\bibitem{ish}C.J. Isham, in Integrable Systems, Quantum Groups
and Quantum Field Theories ed. L.A. Ibort and M.A.
Rodriguez (Kluwer, Dordrecht 1993), gr-qc/9210011.
\bibitem{ash}A. Ashtekar, Gravity and the quantum, New J. Phys. 7, 198 (2005).
\bibitem{and}E. Anderson, The Problem of Time in Quantum Gravity,
in Classical and Quantum Gravity: Theory, Analysis
and Applications, Ed. V. R. Frignanni (Nova, New York 2012), arXiv:1009.2157v3.
\bibitem{mer}Z. Merali,Theoretical physics: The origins of space and time, Nature 500, 516-519 (2013).
\bibitem{marletto}C. Marletto, V. Verdal. Evolution without evolution, and
without ambiguities,    Phys. Rev. D 95, 043510 (2017).
\bibitem{barbour}J. Barbour, The End of Time: The Next Revolution in Physics,
Oxford University Press, (1999)
\bibitem{pag}D.N. Page and W.K. Wootters, Evolution without evolution: Dynamics described by stationary observables Phys. Rev. D, 27, 2885
(1983).
\bibitem{woo}W.K. Wootters, "Time" replaced by quantum correlations Int. J. Theor. Phys. 23, 701 (1984).
\bibitem{pagg}D.N. Page, Clock time and entropy, in Physical Origins of Time Asymmetry, eds. J.J. Halliwell, et al., (Cambridge
Univ. Press, 1993), arXiv:gr-qc 9303020.
\bibitem{gam}R. Gambini, R.A. Porto, J. Pullin, and S. Torterolo, Conditional probabilities with Dirac observables and the problem of time in quantum gravity
Phys. Rev. D 79, 041501(R) (2009).
\bibitem{per}A. Peres, Measurement of time by quantum clocks, Am. J. Phys. 48, 552 (1980).
\bibitem{rov}C. Rovelli, Relational quantum mechanics, Int. J. of Theor. Phys. 35, 1637 (1996),
arXiv:quant-ph 9609002v2.
\bibitem{more}Ekaterina Moreva, Giorgio Brida, Marco Gramegna, Vittorio Giovannetti, Lorenzo Maccone, Marco Genovese, Time from quantum entanglement: an experimental illustration,   Phys. Rev. A 89, 052122 (2014).
\bibitem{mal} J. Maldacena, L. Susskind, Cool horizon for entangled black holes, Fortsch.Phys. 61 (2013) 781-811 arXiv:1306.0533v2(2013).
\bibitem{amps}A. Almheiri, D. Marolf, J. Polchinski and J. Sully, Black Holes:
Complementarity
or Firewalls? JHEP02(2013)062 arXiv:1207.3123 [hep-th].
\bibitem{old}D. N Page, Average entropy of a subsystem, Phys. Rev. Lett. 71, 1291 (1993), arXiv:gr-
qc/9305007; D. N. Page, Black hole information, arXiv:hep-th/9305040 (1993)
\bibitem{mon1}B. M. Terhal, Is entanglement monogamous?,IBM Journal of Research and Development 48, (2004)  arXiv:quant-ph/0307120 (2003).
\bibitem{mon2}M. Koashi and A. Winter, Monogamy of quantum entanglement and other correlations, Phys.
Rev. A 69, 022309 (2004), arXiv:quant-ph /0310037.
\bibitem{fire11} A. Almheiri, D. Marolf, J. Polchinski, D. Stanford, and J. Sully, An apologia for firewalls,
arXiv:1304.6483 (2013).
\bibitem{B1}W. Israel, Thermofield dynamics of black holes, Phys. Lett. A 57, 107 (1976).
\bibitem{B2}J. M. Maldacena, Eternal black holes in anti-de Sitter, JHEP 0304, 021 (2003)
[hep-th/0106112].
\bibitem{preskill}S. Lloyd, J. Preskill, Unitarity of black hole evaporation in the final-state
projection models, JHEP 08 126 (2014)

\bibitem{leny}L. Susskind, Singularities, firewalls, complementarity. 
arXiv:1208.3445v1 [hep-th] 16 Aug 2012
\bibitem{bosso}R. Bousso,Violations of the equivalence principle by a non-locally reconstructed vacuum
at the black hole horizon, Phys. Rev. Lett. 112, 041102 (2014).
\bibitem{non1}B. Czech, J. L. Karczmarek, F. Nogueira and M. Van Raamsdonk, The Gravity
Dual of a Density Matrix, Class. Quant. Grav. 29, 155009 (2012) arXiv:1204.1330
[hep-th].
\bibitem{non2}B. Czech, J. L. Karczmarek, F. Nogueira and M. Van Raamsdonk, Rindler Quantum
Gravity, Class. Quantum Grav. 29 235025 (2012), arXiv:1206.1323  [hep-th].
\bibitem{non3}S. D. Mathur, Black Holes and Beyond,Annals of Physics,327,11 (2012) arXiv:1205.0776 [hep-th].

\end{thebibliography}
\end{document}